\pdfoutput=1
\documentclass[twocolumn,showpacs,floatfix,prl]{revtex4}

\usepackage[final]{graphicx}
\DeclareGraphicsExtensions{.pdf}
\usepackage{amsmath,amssymb,bm}
\usepackage{float}

\begin{document}

\title{Topological invariant in three-dimensional band insulators with disorder}

\author{H.-M. Guo}
\affiliation{Department of Physics, Capital Normal University,
Beijing, 100048, China}
 \affiliation{Department of Physics and
Astronomy, University of British Columbia,Vancouver, BC, Canada V6T
1Z1}

\begin{abstract}
Topological insulators in three dimensions are characterized by a
$Z_2$-valued topological invariant, which consists of a strong index and three
weak indices. In the presence of disorder, only the strong index
survives. This paper studies the topological invariant in disordered
three-dimensional system by viewing it as a super-cell of an
infinite periodic system. As an application of this method we show that
the strong index becomes non-trivial when strong enough disorder is
introduced into a trivial insulator with spin-orbit coupling, realizing a strong topological Anderson insulator. We also numerically extract the
gap range and determine the phase boundaries of this topological
phase, which fits well with those obtained from self-consistent Born
approximation (SCBA) and the transport calculations.
\end{abstract}

\maketitle

Time reversal invariant band insulators of non-interacting electrons
are basically divided into two classes: the ordinary insulator and
the topological insulator \cite{sczhang1, moore1, hasan1, sczhang2}.
The latter is a novel phase of quantum matter. It has an insulating
bulk gap and gapless edge or surface states. These gapless states
are topologically protected and are immune to non-magnetic disorder.
Recently another kind of non-trivial quantum phase termed
topological Anderson insulator (TAI) has been predicted to exist in two dimensions
(2D)\cite{shen, sun, groth} and three dimensions (3D) \cite{guo1}, which
makes the situation more interesting. In the TAI phase, remarkably, the topologically protected gapless states emerge due to disorder.

For systems without disorder, the topological phases can be characterized
by studying the gapless states as obtained e.g.\ from diagonalizing the Hamiltonian in
a geometry with edges or surfaces. They can also be characterized by
the topological invariants calculated from the bulk Hamiltonian,
which have been well studied in recent literature \cite{moore2,
fu1}. However in the disordered systems, gapless modes alone cannot unambiguously identify the topological phases because they may be
localized in space. Instead, the transport properties are usually
used to find these extended topologically protected modes. Similarly
we may also use the topological invariant to characterize the
topological phases induced by disorder. A question naturally
arises: how to calculate the topological invariant in the presence
of disorder.

At first glance, it is not obvious how to generalize the present
methods from translation invariant band insulators to the disordered
systems. Let us first recall the integer quantum Hall effect (IQHE)
where the generalization to the case with disorder is well
understood. The topological quantum number in IQHE, which
characterizes the quantized Hall conductivity, is known as the first
Chern number [also refered to as the
Thouless-Kohmoto-Nightingale-den Nijs (TKNN) integer] and is closely
related to Berry's phase.  In the presence of disorder, the TKNN
integers defined for a clean system can be generalized. By
introducing generalized periodic boundary conditions and averaging
over different boundary condition phase, an invariant expression can
be constructed  applicable to the situation where many-body
interaction and substrate disorder are present \cite{niu1,avron}.
Since the boundary condition phases can be transferred to the
Hamiltonian by a unitary transformation, such considerations are
actually equivalent to thinking of the system as a super-cell of an
infinite system. Since the infinite system which is periodic in the
super-cell has translation symmetry, the wave vectors can be well
defined, which in fact correspond to the boundary condition phases.
The advantage of such consideration is that since the band structure
is recovered, one can use the known methods to calculate the
topological invariant for the infinite system. The finite system
under consideration, which is now a super-cell of the infinite
system, shares the same topological properties as the infinite
system. This idea has been used to study the phase transition in the
presence of disorder in 2D quantum spin Hall system (QSHE)
\cite{essin}, where the authors found that a metallic region always
appears between ordinary and topological insulator in spin-orbit
coupled systems with disorder when there is no extra conservation
law (Recently the same result is also obtained via C*-Algebras
\cite{hastings}).

In this paper, we study the topological invariant of the disordered
3D system by viewing it as a super-cell of an infinite system. We
start from a trivial insulator with spin-orbit coupling and find
that when strong enough disorder is introduced into the system,
a gap appears at half filling and the corresponding
topological invariant is non-trivial, confirming the existence of
the disorder-induced non-trivial topological phase.

\begin{figure}
\includegraphics[width=5cm]{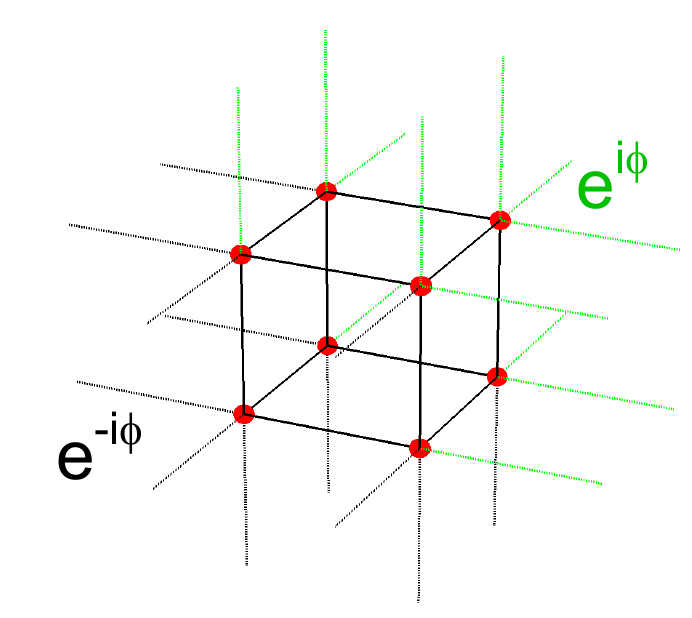}
\caption{(Color online)The cubic lattice on which the Hamiltonian is defined. As an example, the super-cell
 has a size $2^3$ (red closed circles). The sites which have connections with next (previous) super-cell
 obtain phases $e^{i\phi_i}$ ($e^{-i\phi_i}$), where $\phi_i=0$ or $\pi$ and $i=x,y,z$ depends on the TRIM and the direction of the bonds.
 } \label{fig1}
\end{figure}

To be concrete, we consider a model describing itinerant electrons
with spin-orbit coupling on a cubic lattice with  the Hamiltonian in
the momentum space \cite{hosur, rosenberg},

\begin{eqnarray}
    &&H({\bf k})=d_4({\bf k}){\rm I}_{4\times 4}+\nonumber\\
    &&\left(
    \begin{array}{cccc}
        d_0({\bf k})&d_3({\bf k})&0&d_{-}({\bf k})\\
        d_z({\bf k})&-d_0({\bf k})&d_{-}({\bf k})&0\\
        0&d_{+}({\bf k})&d_0({\bf k})&-d_z({\bf k})\\
        d_{+}({\bf k})&0&-d_z({\bf k})&-d_0({\bf k})
    \end{array}
    \right)
    \label{eq:Hk}
\end{eqnarray}
where $d_{\pm}({\bf k})=d_1({\bf k})\pm i d_2({\bf k})$, $d_0({\bf
k})=\epsilon-2t\sum_i \cos k_i$, $d_i({\bf k})=-2\lambda \sin(k_i)$
and $d_4({\bf k})=2\gamma (3-\sum_i \cos k_i)$ ($i=1,2,3$). This
Hamiltonian is a lattice version of the effective low-energy
Hamiltonian describing the physics of insulators in $Bi_2Se_3$
family \cite{sczhang3,hasan2,zxshen,hasan3}. At half filling,
depending on the parameters, the system can be a trivial insulator
or a topological insulator. To simulate the effects of disorder we
consider a random on-site potential $\Sigma_j U_j \Psi_j^+ \Psi_j$,
with $U_j$ uniformly distributed in the range ($-U_0/2,U_0/2$). This
kind of disorder respects time-reversal symmetry. The lattice on
which the Hamiltonian resides is a cubic lattice with a finite size
$L_x\times L_y \times L_z$ (the lattice constant $a=1$). There is no
translation symmetry in the system when disorder is present. However
taking the system as a super-cell of an infinite system, translation
symmetry is recovered. The lattice vector becomes ${\bf a'}_i=L_i$
(the size of the super-cell in $i=x,y,z$ direction) and the
corresponding reciprocal lattice vectors are ${\bf
b}_i=\frac{2\pi}{L_i}$.

 In 3D,
 the topological invariant consists of four $Z_2$ numbers forming an index $(\nu_0;\nu_1 \nu_2 \nu_3)$, which distinguish $16$ topological classes \cite{fu1,guo2}.
 Usually it is a difficult problem to evaluate them for a given band
structure. However in the presence of inversion symmetry, the
problem can be greatly simplified. It has been shown that they can be
determined from the knowledge of the parity $\xi_{2m}({\bm
\Gamma}_i)$ of the 2$m$-th occupied energy band at the 8 time
reversal invariant momenta (TRIM) $\Gamma_i$ that satisfy $\Gamma_i
= \Gamma_i + {\bf G}$. The 8 TRIM can be expressed in terms of
primitive reciprocal lattice vectors as $\Gamma_{i=(n_1 n_2 n_3)} =
(n_1 {\bf b}_1 + n_2 {\bf b}_2 + n_3 {\bf b}_3)/2$, with $n_j =
0,1$. Then $\nu_\alpha$ is determined by the product $(-1)^{\nu_0} =
\prod_{n_j = 0,1} \delta_{n_1 n_2 n_3},$ and $(-1)^{\nu_{i=1,2,3}} =
\prod_{n_{j\ne i} = 0,1; n_i = 1} \delta_{n_1 n_2 n_3}$, where the
parity product for the occupied bands $\delta_i=\prod_{m=1}^{N}
\xi_{2m}(\Gamma_i)$. To take advantage of the simplification, we
only consider disorder configurations with inversion symmetry.
For large enough super-cell such consideration will not change the underlying physics. So
to calculate the topological invariant, we only need to consider the
Hamiltonian at the eight TRIM and they are equivalent to those of a
finite system with boundary conditions which are periodic up to
phases $\phi_x, \phi_y, \phi_z$=$0$ or $\pi$ for boundary sites, as
shown in Fig. \ref{fig1}.

\begin{figure*}
\includegraphics[width=12cm]{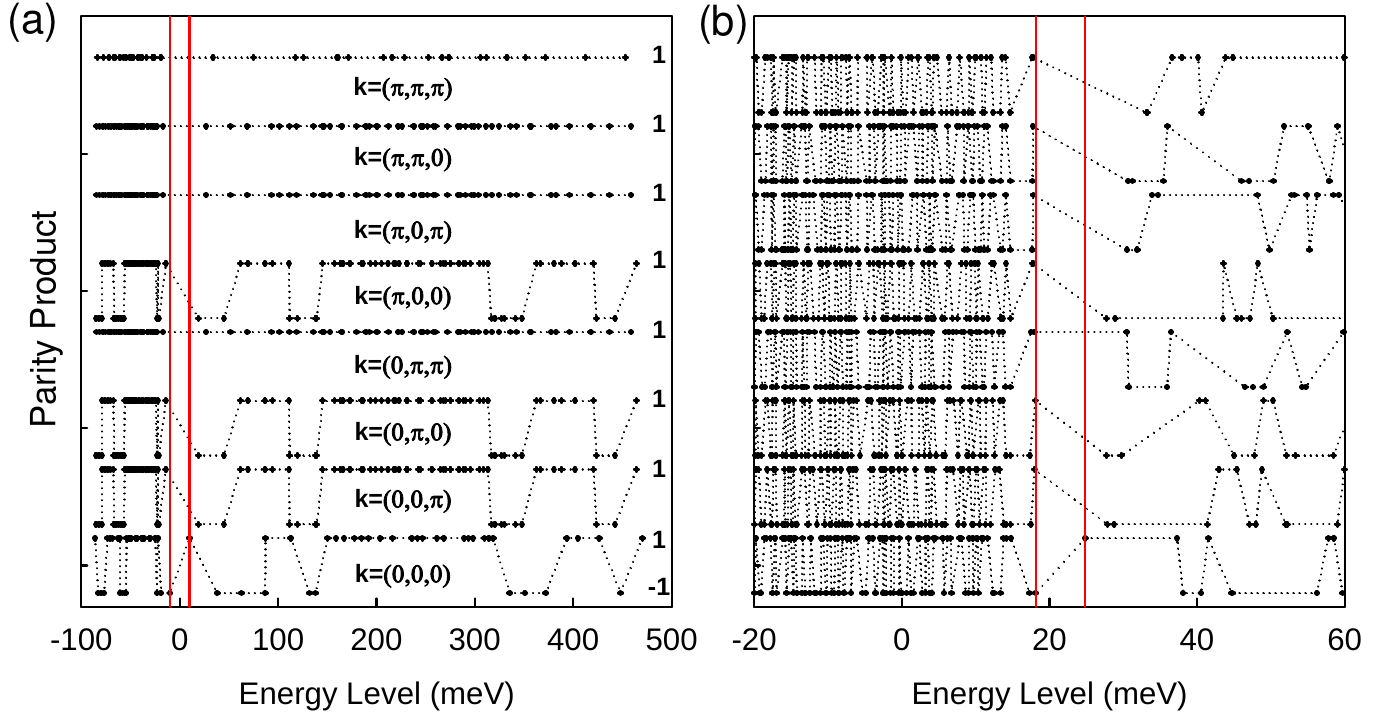}
\caption{(Color online) The parity product for the occupied states
at the eight TRIM for a $8^3$ super-cell without disorder (a) and
with disorder (b). The crosses mark the eigenvalue of the system and
the corresponding parity product for the filling up to this energy
level. Here we use parameters: $t=24$meV, $\lambda=20$meV,
$\gamma=16$ meV and (a) $\epsilon=134$meV, corresponding to $m=-10$
meV, where the system is a $(1;000)$ strong topological insulator;
(b)$\epsilon=145$ meV and $U_0=150$ meV, corresponding to $m=1$ meV,
where the clean system is a trivial insulator. The red lines in both
figures show the gap range which appears at half filling.
 } \label{fig2}
\end{figure*}

To understand what happens when introducing the super-cell, we first
calculate the $Z_2$ topological invariant of a clean $(1;000)$
strong topological insulator. The results are shown in Fig.
\ref{fig2} (a). At half filling, we find that $\delta=-1$ at the
$\Gamma$ point and $\delta=1$ at other TRIM. This result is
consistent with the ordinary band structure calculations and the
reason is explained below. Taking a super-cell means enlarging the
original unit cell in the ${\bf a}_1,{\bf a}_2,{\bf a}_3$ directions
(${\bf a}_1,{\bf a}_2,{\bf a}_3$ are the lattice vectors of the
original lattice). The resulting new Brillouin zone (BZ) is folded
in the corresponding directions and shrunk in size. The occupied
states at the eight TRIM for the new BZ contain those at the eight
TRIM for the original BZ. Though more states which are at other
momenta of the original BZ will reside on the TRIM of the new BZ,
they do not change the parity product for the occupied states. Thus
the product of all eight $\delta$s from the super-cell calculation
still yield the same 'strong' index of the $Z_2$ topological
invariant. However the 'weak' index cannot be obtained from this
method if the super-cell contains an even number of unit cell in the
corresponding direction. As mentioned above, the weak indices
$\nu_1,\nu_2,\nu_3$ are the product of four $\delta$s which are in
the planes ${ k}_1=\pi$,${ k}_2=\pi$, ${k}_3=\pi$ respectively.
Suppose that the super-cell contains an even number of unit cell in
the ${\bf a}_1$ direction, then the ${ k}_1=0$ and ${k}_1=\pi$
planes will collapse onto the ${k}_1'=0$ plane in the new BZ. Thus
the weak index determined by the four TRIM on the $k_1'=\pi$ plane
in the new BZ must be $0$. From another point of view, this is
understandable because the 'weak' indices are related to layered 2D
quantum spin-Hall states and the 3D super-cell naturally fails in
calculating the quantities reflecting the 2D physics. When
considering systems with disorder, the weak indices are eliminated
and only the strong index remains
 robust. The topological invariant for the system with disorder is
the 'strong' index of the $Z_2$ topological invariant for clean 3D
band insulators. In the following, we simply call it the
topological invariant for the 3D systems with disorder.

\begin{figure}
\includegraphics[width=8cm]{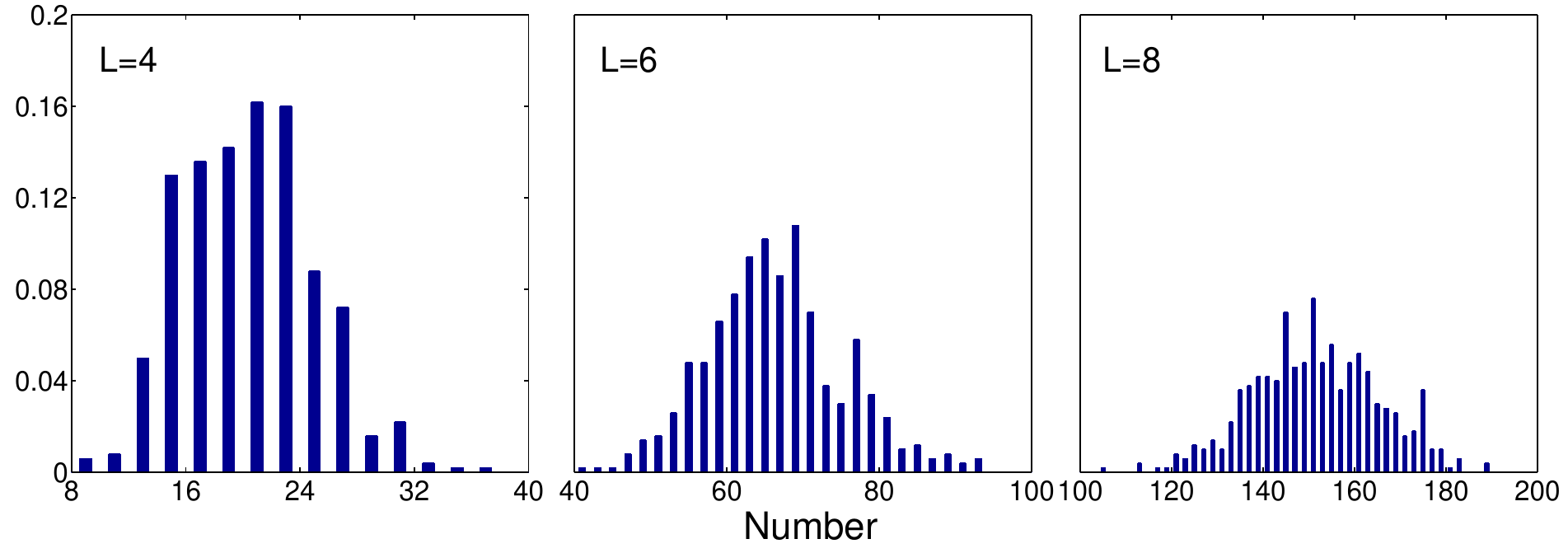}
\caption{(Color online)Distributions of the number of $Z_2$-odd band pairs from 500 disorder realizations.
The bar heights is the fraction of disorder realizations that have a given number of band pairs with $Z_2=1$ (The numbers are all odd here).
The parameters are the same as those in Fig. \ref{fig2} (b).
 } \label{fig3}
\end{figure}

Now we take into account disorder as described above and start from
a trivial insulator. Fig. \ref{fig2} (b) shows the result of such a
calculation on a $8^3$ lattice with $U_0=150$ meV. Generally the
disorder will eliminate all degeneracies except those protected by
time reversal symmetry. So at the TRIM, each eigenvalue of the
Hamiltonian is doubly degenerate. Though the disorder makes the
energy spectrum more continuous, a small gap remains clear at half
filling. Here the gap location is shifted to the high energy, which
is different from the case in the clean system where the gap appears
symmetrically at $E=0$. The parity eigenvalue products $\delta=-1$
at the $\Gamma$ point and $\delta=1$ at other TRIM for half filling
show that the topological invariant of the system is $1$, so the
system will exhibit non-trivial topological properties. We have
extended the above calculation to $500$ different disorder
realizations. In all $500$ disorder realizations, the
 topological invariant is $1$ for half filling, which further confirms the topological phase in the system. Thus through
calculating the topological invariant of the system, we obtain
further confirmation of the topological phase induced by disorder.
This new phase has been termed as 'strong topological Anderson
insulator' (STAI) \cite{guo1}. Due to the non-trivial topological
properties in STAI,  topologically protected surface states will
appear at the surfaces as the case in the 'strong topological
insulator' phase of the clean system, which has already been
verified by transport calculations.

 It is also interesting
to look at the number of $Z_2$-odd band pairs. In 2D QSHE, the
number of $Z_2$-odd Karmers pairs increases linearly with the system
size in the metallic region and grows slowly for the topological
insulator \cite{essin}. Here in 3D, we find a similar result. In
Fig. \ref{fig3}, we show the distribution of the number of $Z_2$-odd
band pairs  at three lattice sizes $4^3$, $6^3$ and $8^3$. We find
that the location of the mean roughly scales with $L^3$ ($L$ is the
lattice size). A detailed analysis on the limited data also shows
that the number growth is somewhat slower than linear. With enough
data and larger sizes, finite-size scaling could be carried out.
However this is not accessible at present due to the limited
computer resources.

\begin{figure}
\includegraphics[width=8cm]{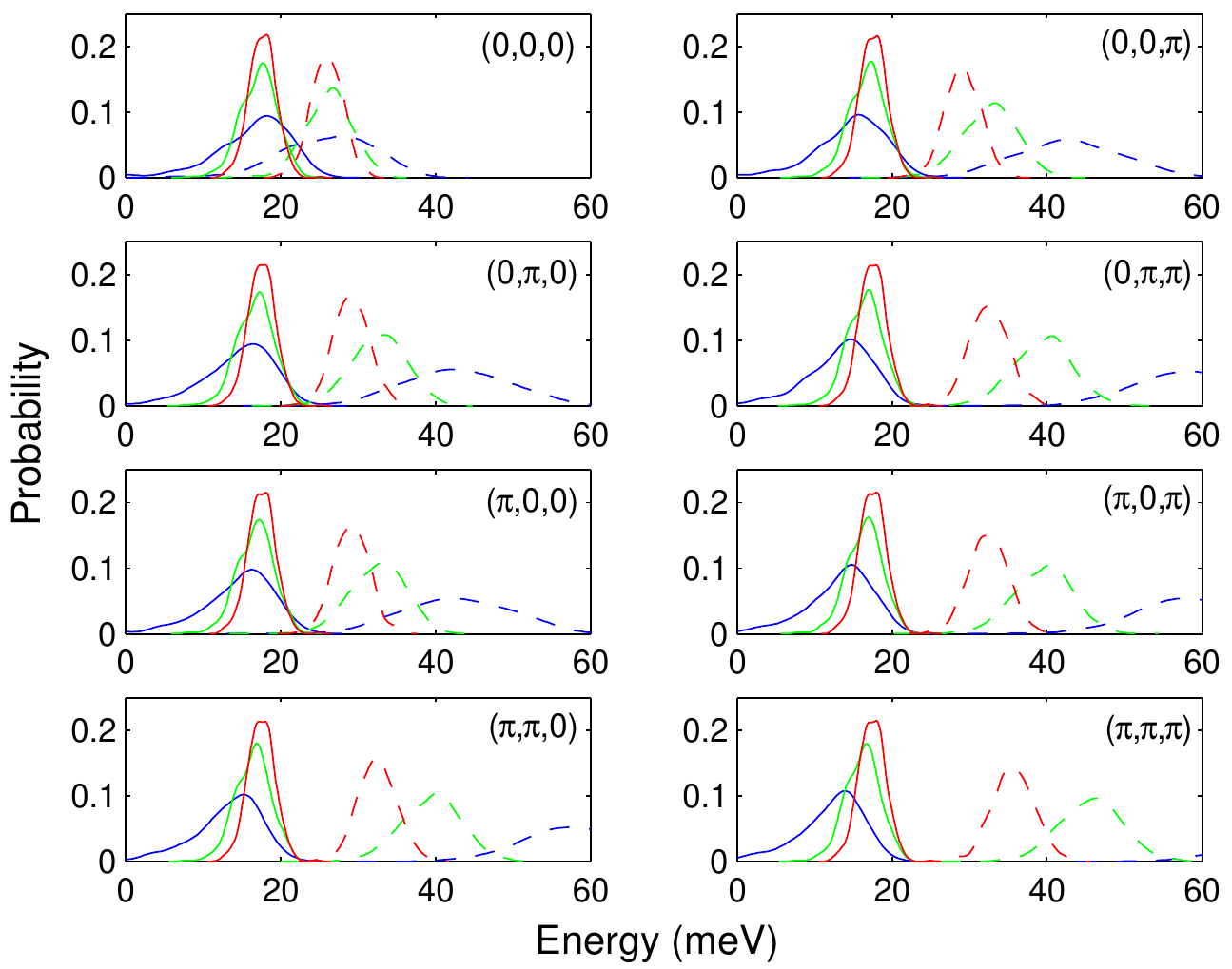}
\caption{(Color online) Distributions of energy levels at half filling and half plus one filling
from 500 disorder realizations. The system sizes are $4^3$(blue),$6^3$(green) and $8^3$(red). The parameters are the same as those in Fig. \ref{fig2} (b).
 } \label{fig4}
\end{figure}

For different disorder realizations, though there still exits a gap
at half filling, its position on the energy axis changes randomly.
We extract the energy levels for the most energetic electron at half
filling and the one just above it and show their distributions in
Fig. \ref{fig4}. The curves in Fig. \ref{fig4} show Gaussian-like
distributions. The distributions for half filling and half plus one
at some TRIM have no overlaps while on the other TRIM have overlaps.
This means there is no 'true' gap range even for the eight TRIM.
However the overlaps are already very small for the lattice size of
$8^3$. We also performed the same calculations on lattice sizes of
$4^3$ and $6^3$. With the increasing lattice size the widths of the
distributions decrease. We therefore attribute the overlaps for the
present lattice size to finite size effect and expect them to
diminish for larger lattice sizes.

\begin{figure}
\includegraphics[width=8cm]{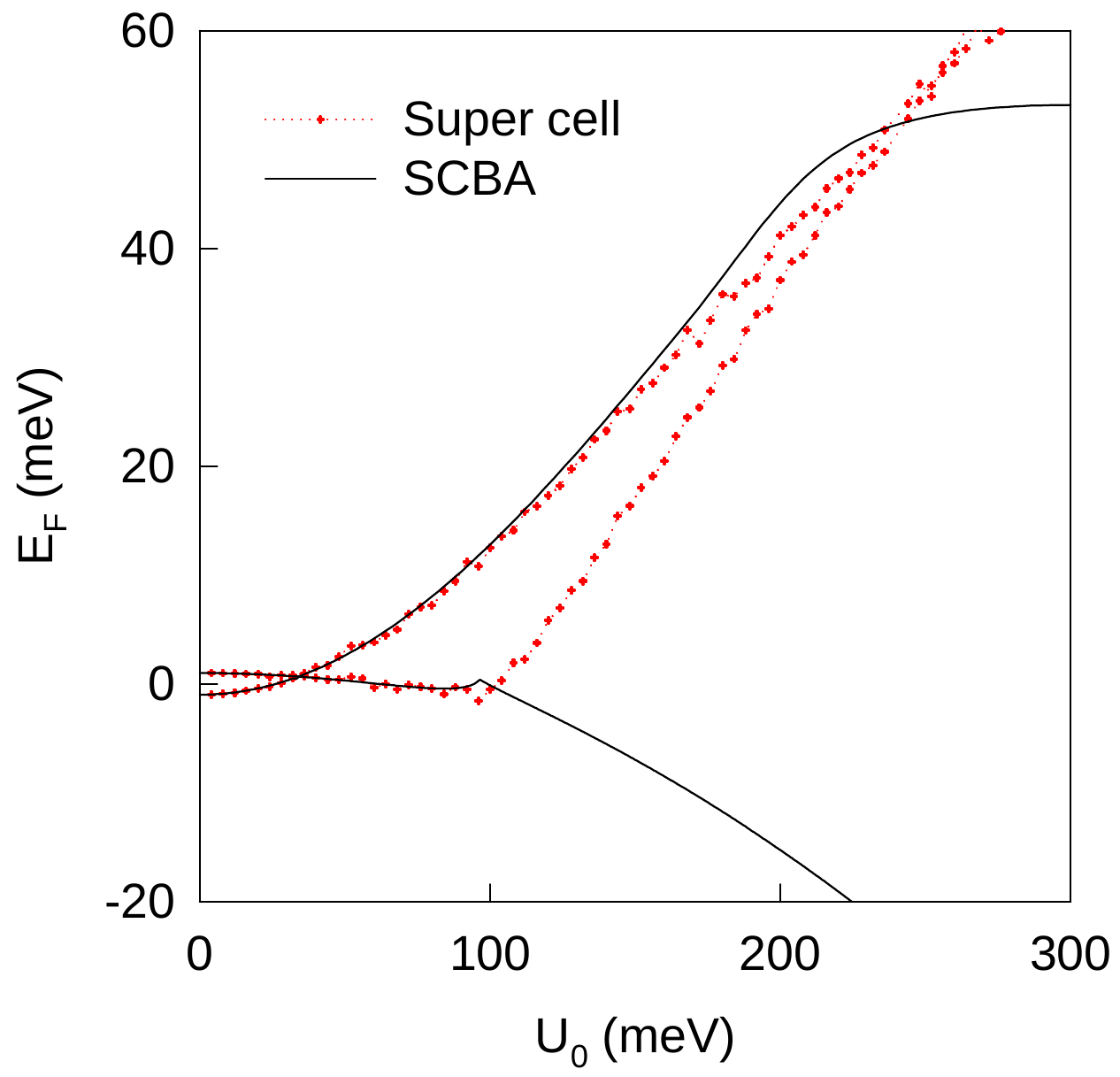}
\caption{(Color online) Phase boundaries for the STAI phase in the $U_0-E_F$ plane obtained from the super-cell calcuations.  The red crosses represent the peaks in energy level distributions
 for half and half plus one fillings and are obtained from calculations of 500 disorder realizations on $6^3$ system. The phase boundaries from
 the SCBA are also shown for comparison.
 } \label{fig5}
\end{figure}

For the parameters used the physical gap size is determined by gap at ${\bf k}=(0,0,0)$ TRIM. We can obtain the gap value from the peaks of
the distributions and this value will approximate the one for the
larger lattice size. We have determined the phase diagram in the $U_0-E_F$
plane from calculations on a $6^3$ system and the result is shown in
Fig. \ref{fig5}. The range of STAI in the $U_0-E_F$ plane has also been
obtained from conductivity calculations and the SCBA, where the
weak-disorder boundary fits well with each other while the
strong-disorder boundary does not \cite{guo1}. It was found that the
weak-disorder boundary marks the crossing of a band edge and the
strong-disorder boundary marks the crossing of a mobility edge, and
The SCBA does not work for the strong-disorder boundary
\cite{groth}.

The result in Fig. \ref{fig5} shows that the phase boundaries
obtained from the present calculation fits well for both the
weak-disorder and strong-disorder phase boundaries determined from
the conductivity calculation \cite{guo1}. The range of STAI phase
should be determined by the gap between the two edges containing
extended states. We believe that we have extracted such a gap. The
reason is that the localized states are closely related to the
disorder realizations and can be removed by an average over disorder
realizations. In the phase diagram, increasing the strength of
disorder, the system first undergoes a sharp phase transition, which
happens at a critical disorder strength where the gap closes. Then
the system enters a stable topological phase for a wide range of
disorder strength. Finally the system experiences another phase
transition to a diffusive metal. Close to the latter transition, the
finite system under consideration is in a mixed phase, where some
fraction of disorder realizations yield a non-trivial topological
invariant.


In conclusion, we have generalized the method of calculating the
topological invariant in disordered 2D QSHE to 3D disordered
systems. In this method, the finite 3D system is viewed as a
super-cell of a large lattice with well defined wave vector, which
allows us to directly use the definition of topological invariants for
clean band insulators. The obtained topological invariant can be
thought of as describing the finite system with disorder. Using
the inversion symmetry preserving disorder configurations, we carried out explicit
calculations on a model Hamiltonian (describing the physics of
insulators in $Bi_2Se_3$ family) with on-site disorder. We found that
the topological invariant for a system that is a trivial insulator in the absence of disorder can become nontrivial when strong
enough disorder is introduced. This result confirms
the existence of the strong topological Anderson insulator, a topological phase in three space dimensions whose existence is fundamentally dependent on disorder \cite{guo1}.

As additional application of this method we counted the number of $Z_2$-odd band
pairs. Though this number varies with the disorder realization, the total
number is always odd when the system is in the STAI phase and it
scales roughly linearly with the number of the lattice sites. Finally, we
obtained the phase boundaries for the STAI phase by extracting the gap
range, which fits well with the known results.

In the presence of on-site disorder, the diagonal elements of the
Hamiltonian matrix are random, drawn from a statistical distribution. The Hamiltonian can be regarded as a random matrix which
can be studied in the framework of the random matrix theory (RMT).
The RMT has already been applied to IQHE  and disordered superconductors
\cite{rmt1,rmt2,rmt3}. We expect some insights into the present
problem from RMT, which we leave to future work.

\emph{Acknowledgment}.--- Authors are indebted to I. Garate, M. Franz, G.
Refael, G. Rosenberg, C. Weeks and M. Vazifeh  for stimulating
discussions. Support for this work came from NSERC, CIfAR and The
China Scholarship Council.

\end{document}